\documentclass[apj,onecolumn,numberedappendix]{emulateapj}
\usepackage{txfonts}
\usepackage{bm}

\newcommand\F[4]{\ensuremath{\mbox{}_2F_1\!\left(#1,#2;#3;#4\right)}}
\renewcommand\email\texttt

\begin{document} 

\slugcomment{\sc to appear in \it the Astronomical Journal}
\shorttitle{\sc
Galaxy Models with Tangentially Anisotropic Velocity Distributions}
\shortauthors{\sc J.~An \& N.~W.~Evans}

\title{Galaxy Models with Tangentially Anisotropic Velocity
Distributions}

\author{Jin H. An\altaffilmark{1} and N. Wyn Evans}
\affil{Institute of Astronomy, University of Cambridge,
Madingley Road, Cambridge CB3 0HA, UK;\\
\email{jinan@space.mit.edu, nwe@ast.cam.ac.uk}}
\altaffiltext{1}{Current Address:
MIT Kavli Institute for Astrophysics \& Space Research,
Massachusetts Institute of Technology,
77 Massachusetts Avenue, Cambridge, MA 02139, USA}

\begin{abstract}
This paper provides two families of flexible and simple galaxy models.
Many representatives of these families possess important
cosmological cusps, with the density behaving like $r^{-1}$,
$r^{-4/3}$, or $r^{-3/2}$ at small radii. The density falls off between
$r^{-3}$ and $r^{-5}$ at large radii. We provide analytic and
anisotropic distribution functions for all the models. Unlike many
existing methods, our algorithm can yield tangentially anisotropic
velocity dispersions in the outer parts, and so is useful for modeling
populations of satellite galaxies and substructure in host galaxy
halos. As an application, we demonstrate the degeneracy between mass
and anisotropy for the satellite galaxy population of the Milky Way.
This can introduce a factor of $\sim$3 uncertainty in the mass of the
Milky Way as inferred from the kinematics of the satellite population.
\end{abstract}


\section{Introduction}

One of Eddington's famous discoveries is that the isotropic
distribution function (DF) of a spherical stellar system can be
calculated from the density using an Abel transform pair
\citep[p.\ 237]{Ed16,BT87}:
\begin{equation}
f(E)=\frac1{\sqrt 8\pi^2}
\left(\int_0^E\frac{d^2\rho}{d\psi^2}\frac{d\psi}{\sqrt{E-\psi}}
+\frac 1{\sqrt E}\left.\frac{d\rho}{d\psi}\right|_{\psi=0}\right).
\label{eq:eddington}
\end{equation}
Here, $E=\psi-v^2/2$ is the binding energy per unit mass, and $\psi$ is
the relative potential.

However, galaxy halos produced in cosmological simulations are held up
by anisotropic velocity dispersions \citep[see e.g.,][]{HM05}.
The recovery of an
anisotropic DF for a spherical system is a more difficult problem
\citep{HQ93}. Now, the steady state DF can depend not only on $E$ but
also on the magnitude of the specific angular momentum $L=rv_\mathrm T$
through Jeans' theorem. Here $v_\mathrm T$ is the two-dimensional
velocity component projected on the tangential plane.

At least in the comparison of observational data with models
\citep[e.g.,][]{vdM00,Wi04,TK04}, a widely used \textit{Ansatz}
for the form of DF is
\begin{equation}
f(E,L)=L^{-2\beta}f_E(E),
\label{eq:ansatz}
\end{equation}
where $f_E(E)$ is a function of the binding energy alone. It is easy
to show that the models generated by this DF exhibit the property that
\begin{equation}
\beta=1-\frac{\langle v_\mathrm T^2\rangle}{2\langle v_r^2\rangle}.
\label{eq:beta}
\end{equation}
Here, $\langle v_r^2\rangle$ and $\langle v_\mathrm T^2\rangle$ are
the radial and the tangential velocity second moments. The general
inversion formula for the unknown function $f_E(E)$ \citep[see
also Appendix~\ref{sec:con}]{Cu91,Ko96,WE99} is no more difficult
than Eddington's formula (eq.~\ref{eq:eddington}).

We note that it is common practice to characterize the velocity
anisotropy of a spherically symmetric stellar system by the parameter
$\beta$ defined in equation (\ref{eq:beta}), which is sometimes
referred to as the anisotropy parameter \citep{BT87}. If
$0<\beta\le1$, the velocity dispersion ellipsoid is a prolate spheroid
with the major axis along the radial direction (radially anisotropic),
while it is an oblate spheroid with the tangential plane being the
plane of symmetry (tangentially anisotropic) if $\beta<0$. In
particular, $\beta=1$ and $\beta=-\infty$ indicate that every star is
in a radial or circular orbit, respectively. On the other hand,
systems with isotropic velocity dispersions have $\beta=0$. In
general, the anisotropy parameter varies radially, but for the system
with a DF of the form of equation (\ref{eq:ansatz}), it is constant
everywhere.

Despite the attractiveness of a simple inversion of the DF in
equation (\ref{eq:ansatz}), the assumption of
uniform anisotropy is not always desirable. For example, it is useful
to be able to build models that are isotropic in the center and
radially anisotropic in the outer parts, as seems to be the case for
dark matter halos \citep{HM05}. DFs that are tangentially anisotropic
in the outer parts are helpful for understanding the substructure and
the satellite galaxy populations in dark matter halos.

In this paper we study algorithms for building galaxies with varying
anisotropy, using sums of DFs of the form of equation(\ref{eq:ansatz}).
\S~\ref{sec:gis}
introduces a new family of cusped halo models, the generalized
isochrones, and \S~\ref{sec:gp} examines the generalized Plummer models.
Every member of the family has analytic potential-density pairs,
velocity dispersions, and DFs. Specific examples are provided with the
cosmologically important $r^{-1}$, $r^{-3/2}$, and $r^{-4/3}$ density
cusps. \S~\ref{sec:app} discusses a specific
application of our theoretical models
to the problem of estimating the mass of a host galaxy from the motion
of its satellites. Finally, \S~\ref{sec:conc} concludes with a brief comparison
between our methods of producing varying-anisotropy DFs and other
suggestions in the literature \citep[e.g.,][]{Os79,Me85,Cu91,Ge91}.

\section{The Generalized Isochrone Spheres}
\label{sec:gis}

\begin{figure}
\epsscale{0.5}
\plotone{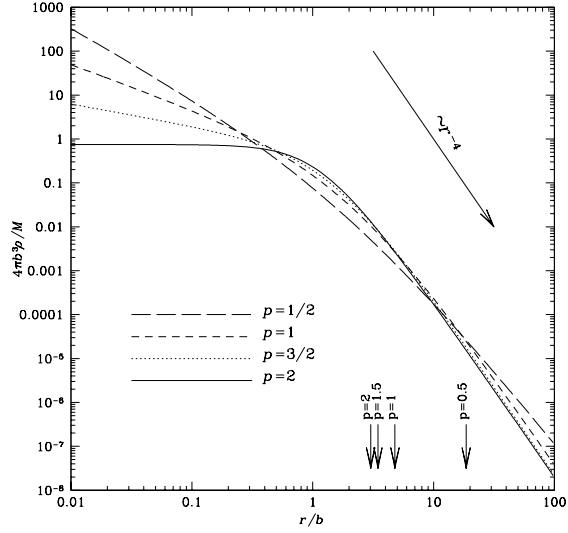}
\caption{\label{fig:geniso} Density profile of the generalized
isochrone models. The classical isochrone ($p=2$) is cored, whereas
all other models are cusped with $\rho\sim r^{-(p-2)}$ at small radii.
The small arrows indicate the half-mass radius.}
\end{figure}

The isochrone sphere has the potential-density pair
\citep{He59,ELS,ELZ}
\begin{equation}
\psi=\frac{GM}{b+s}\,;\qquad
\rho=\frac M{4\pi}\frac{(2s+b)b}{s^3(b+s)^2},
\label{eq:iso}
\end{equation}
where $M$ is the total mass of the system, $s=\sqrt{b^2+r^2}$, and $b$
is a constant defining the scale length. Like the \citet{Pl11} model,
the density is finite and well behaved at the center. Unlike the
Plummer model, the density falls off as $r^{-4}$ at large radii, which
is a better fit to elliptical galaxies than the $r^{-5}$ fall-off of the
Plummer model. There are many available studies regarding its DFs
\citep[e.g.,][]{Ge91,BL97}, some of which are essentially numerical in
nature.

Inspired by this, let us consider the potential-density pair
\begin{equation}
\psi=\frac{GM}{b+s}\,;\qquad
\rho=\frac M{4\pi}\frac{2br^p+b^p(1+p)(b+s)}{r^{2-p}s^{2p-1}(b+s)^3},
\label{eq:gis}
\end{equation}
where
\begin{equation}
s=\left(b^p+r^p\right)^{1/p}.
\end{equation}
Here, $p>0$ is a new parameter used for the generalization of the
isochrone sphere (eq.~\ref{eq:iso}). The condition that the density
decreases monotonically with increasing radius restricts $p\le2$, with
the limit ($p=2$) being the cored classical isochrone sphere. In fact,
all the halo models with $0<p<2$ are cusped. In particular, the
density behaves as $\rho\sim r^{-(2-p)}$ as $r\rightarrow0$. On the
other hand, we find that $\rho\sim r^{-q}$ where $q=\min(4,p+3)$ as
$r\rightarrow\infty$. The density profiles of some members of this
family are shown in Figure~\ref{fig:geniso}. We note that the case
$p=1$, for which $\psi=GM/(2b+r)$, is actually recognized as the
\citet{He90} model, albeit with a slightly unusual identification for the
scale length ($a=2b$). Models with $p=1/2$ and $p=2/3$ possess the
cosmological cusps suggested by \citet{Mo98} and \citet{EC97}.

The enclosed mass and the circular speed are found to be
\begin{equation}
M_r=-\frac{r^2}G\frac{d\psi}{dr}
=M\left(1+\frac{b^p}{r^p}\right)^{1/p-1}
\left[\left(1+\frac{b^p}{r^p}\right)^{1/p}+\frac br\right]^{-2}
\,;\qquad
v_\mathrm c^2=\frac{GM_r}r=\frac{GMr^p}{s^{p-1}(b+s)^2}.
\end{equation}
The potential $\psi(r)$ of the generalized isochrone sphere is simple
enough that the inverse, $r(\psi)$, can be easily found (with $G=M=b=1$
without any loss of generality):
\begin{equation}
r(\psi)=\frac{\left[(1-\psi)^p-\psi^p\right]^{1/p}}\psi.
\end{equation}
Using this, in principle, one can construct infinitely many different
expressions for $\rho(\psi,r)$, each of which can be inverted to yield
the DF by means of integral transforms \citep{Ly62}. However, for most
cases this procedure is analytically intractable. Rather, it is often
more productive to look for a specific form of $\rho(\psi,r)$ that is
associated with a plausible \textit{Ansatz} for the DF, whose inversion
reduces to elementary integrals.

\subsection{Two Component Distribution Functions}

One such possible ($r$,$\psi$)-split with an analytically tractable
inversion is given by
\begin{equation}
\rho=
\frac{p+1}{4\pi}\,r^{p-2}\psi^{2p+1}(1-\psi)^{1-2p}
+\frac1{2\pi}\,r^{2p-2}\psi^{2p+2}(1-\psi)^{1-2p}.
\label{eq:expone}
\end{equation}
We note that a DF of the form of equation (\ref{eq:ansatz}) can be
inverted from an ($r$,$\psi$)-split that is a monomial of $r$, with its
power index linearly related to $\beta$. By having the
($r$,$\psi$)-split given by a linear combination of two monomials of
$r$, the inversion of equation (\ref{eq:expone}) can be achieved by
simply extending the known procedure, while the system is no longer
restricted to have uniform velocity anisotropy. Inspired by this, let
us suppose that the DF is of the form
\begin{equation}
f(E,L)=L^{p-2}f_1(E)+L^{2p-2}f_2(E),
\label{eq:diffiso}
\end{equation}
where $f_1(E)$ and $f_2(E)$ are functions of $E$ to be determined.
Then, the density is found to be
\begin{equation}
\rho=
r^{p-2}\,\frac{2^{p/2+1/2}\pi^{3/2}\Gamma(p/2)}{\Gamma(p/2+1/2)}
\int_0^\psi\!dE\,(\psi-E)^{p/2-1/2}f_1(E)
+
r^{2p-2}\,\frac{2^{p+1/2}\pi^{3/2}\Gamma(p)}{\Gamma(p+1/2)}
\int_0^\psi\!dE\,(\psi-E)^{p-1/2}f_2(E),
\end{equation}
where $\Gamma(x)$ is the gamma function.
By comparing this to equation (\ref{eq:expone}), a possible DF may be
found by inverting the integral equations
\begin{eqnarray}
\int_0^\psi\!dE\,(\psi-E)^{p/2-1/2}f_1(E)
&=&\frac{(p+1)\Gamma(p/2+1/2)}{2^{p/2+5/2}\pi^{5/2}\Gamma(p/2)}\
\psi^{2p+1}(1-\psi)^{1-2p}
\,;\nonumber\\
\int_0^\psi\!dE\,(\psi-E)^{p-1/2}f_2(E)
&=&\frac{\Gamma(p+1/2)}{2^{p+3/2}\pi^{5/2}\Gamma(p)}\
\psi^{2p+2}(1-\psi)^{1-2p}.
\end{eqnarray}
The inversion procedure is essentially identical to the one that leads
to equations~(\ref{eq:disint}) and (\ref{eq:dishalf}). After this
inversion, we find that $f_1(E)$ and $f_2(E)$ are
\begin{eqnarray}
f_1(E)&=&
\frac{\Gamma(2p+3)}{2^{p/2+7/2}\pi^{5/2}\Gamma(p/2)\Gamma(3p/2+3/2)}\
E^{(3p+1)/2}\F{2p-1}{2p+2}{\frac{3p+3}2}E
\,;\nonumber\\
f_2(E)&=&
\frac{\Gamma(2p+3)}{2^{p+3/2}\pi^{5/2}\Gamma(p)\Gamma(p+5/2)}\
E^{p+3/2}\F{2p-1}{2p+3}{p+\frac52}E,
\label{eq:dfgeniso}
\end{eqnarray}
both of which are always nonnegative for all accessible $E$-values
($0\le E\le\psi\le1/2$). While we have given the result in complete
generality in terms of hypergeometric functions $\F{a}{b}{c}{x}$,
the DFs reduce to entirely elementary and analytic functions if $p=1/2$,
1, or 2. In fact, if $p=1/2$, both hypergeometric functions are just
unity, whereas if $p$ is an integer, both reduce to elementary
functions. If $p$ is a half integer other than $1/2$, only $f_2$
reduces to an elementary function.

Of particular interest is the case in which $p=1/2$, for which the model
possesses a central cusp like $\rho \sim r^{-3/2}$, as found by
cosmological simulations \citep{Mo98}. The corresponding DF is very
simple, namely,
\begin{equation}
f(E,L)=\frac3{4\pi^3}\frac{E^2}L
+\frac{2^{5/4}\cdot3}{5\pi^{5/2}\Gamma(1/4)^2}\frac{E^{5/4}}{L^{3/2}}.
\end{equation}
If $p=1$ (the Hernquist model with a rescaled scale length),
\begin{equation}
f(E,L)=\frac1{2^{7/2}\pi^3(1-E)^2}
\left[\frac{3\arcsin\!\sqrt E}{\sqrt{1-E}}
-\left(1-2E\right)\left(3+8E-8E^2\right)\sqrt E\right]
+\frac1{4\pi^3L}\frac{E^2(3-2E)}{(1-E)^2},
\end{equation}
or if $p=2$ (the classical isochrone sphere);
\begin{eqnarray}
f(E,L)&=&
\frac{3L^2}{2^{17/2}\pi^3(1-E)^5}
\left[(40E^2-24E+5)\,\frac{15\arcsin\!\sqrt E}{\sqrt{1-E}}
-\left(75-310E+400E^2-928E^3+576E^4-128E^5\right)\sqrt E\right]
\nonumber\\&+&
\frac3{2^{15/2}\pi^3(1-E)^4}
\left[(16E^2-12E+3)\,\frac{15\arcsin\!\sqrt E}{\sqrt{1-E}}
-\left(45-150E+144E^2-208E^3+64E^4\right)\sqrt E\right].
\end{eqnarray}
Both of these DFs are no more complicated than the isotropic DFs deduced
by \citet{He90} and \citet{He60}, respectively.

\subsubsection{Kinematics}

\begin{figure}
\epsscale{0.4}
\plotone{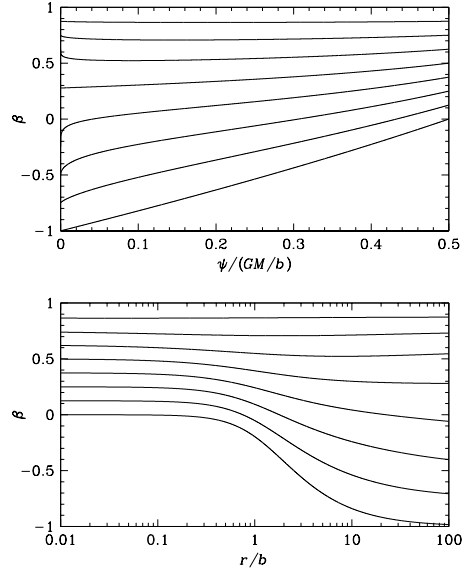}
\caption{\label{fig:anisoc} Behavior of the anisotropy parameter
generated by the DFs (eq.~\ref{eq:diffiso}) for the
generalized isochrone spheres. From top to bottom, the parameter $p$
varies from 1/4 to 2 with increments of 1/4. Note that the last curve,
$p=2$, corresponds to the classical cored isochrone sphere.}
\end{figure}

To find the velocity dispersion of these models, we exploit the
splitting $\rho=\rho_1+\rho_2$, where
\begin{equation}
\rho_1=
\int L^{p-2}f_1(E)\,d^3\!\bm v
=\frac{(p+1)r^{p-2}\psi^{2p+1}(1-\psi)^{1-2p}}{4\pi}
\,;\qquad
\rho_2=
\int L^{2p-2}f_2(E)\,d^3\!\bm v
=\frac{r^{2p-2}\psi^{2p+2}(1-\psi)^{1-2p}}{2\pi}.
\end{equation}
Then,
\begin{equation}
\rho\langle v_r^2\rangle
=\frac{\rho_1\psi}{2(p+1)}V_1 +\frac{\rho_2\psi}{2p+3}V_2\,;\qquad
\rho\langle v_\mathrm T^2\rangle
=\frac{p\rho_1\psi}{2(p+1)}V_1 +\frac{2p\rho_2\psi}{2p+3}V_2,
\end{equation}
where
\begin{equation}
V_i=(1-\psi)\,\F{3+i}{1}{2p+2+i}{\psi}
=\F{2p-1}{1}{2p+2+i}{-\frac\psi{1-\psi}}.
\end{equation}
The anisotropy parameter
\begin{equation}
\beta=1-\frac p2\,
\left[\frac{(2p+3)V_1+8r^p\psi V_2}{(2p+3)V_1+4r^p\psi V_2}\right]
\label{eq:gicvi}
\end{equation}
is now no longer constant. Figure~\ref{fig:anisoc} shows the behavior
of $\beta$ for some of these models. The curves can be understood on
examination of equation (\ref{eq:gicvi}). This reveals that whereas
$\beta=1-(p/2)$ at $r=0$ for any $p$,
$\lim_{r\rightarrow\infty}\beta=1-p$ if $p>1$, or
$\lim_{r\rightarrow\infty}\beta=1-(p/2)$ if $p<1$. At the critical
value $p=1$, we find that the anisotropy parameter monotonically
decreases from $\beta=1/2$ at $r=0$ to $\beta\rightarrow5/18$ as
$r\rightarrow\infty$. Hence, the DFs discussed in this section build
the generalized isochrone spheres with either decreasing $\beta$ in
the inner region and increasing $\beta$ in the outer region ($p<1$), or
decreasing $\beta$ throughout ($p\ge1$). The particular case of the
latter also includes the Hernquist model ($p=1$) and the classical
isochrone sphere ($p=2$).

The simplest case is again $p=1/2$, which leads to
\begin{equation}
\langle v_r^2\rangle
=\frac{4\rho_1+3\rho_2}{\rho_1+\rho_2}\frac\psi{12}
=\frac{2+3r^{1/2}+r}{6+10r^{1/2}+3r}\,\psi;\qquad
\langle v_\mathrm T^2\rangle
=\frac{2\rho_1+3\rho_2}{\rho_1+\rho_2}\frac\psi{12}
=\frac{2+4r^{1/2}+r}{6+10r^{1/2}+3r}\,\frac\psi2.
\end{equation}
Then we find that the virial theorem holds locally for this case:
\begin{equation}
\langle v_r^2\rangle+\langle v_\mathrm T^2\rangle
=\frac{6\rho_1+6\rho_2}{\rho_1+\rho_2}\frac\psi{12}=\frac\psi2,
\end{equation}
and the anisotropy parameter
\begin{equation}
\beta=\frac14\,\frac{6+8r^{1/2}+3r}{2+3r^{1/2}+r}
=\frac{3(r-2)^2-r^{1/2}(r+2)}{4(r-4)(r-1)}
\end{equation}
varies from $\beta=3/4$ at $r=0$, to $\beta=1/\sqrt 2$ at $r=2$, and back
to $\beta\rightarrow3/4$ as $r\rightarrow\infty$. This model in fact
corresponds to one of the generalized hypervirial models studied by
\citet{AE05a}.
That is, $\psi=[1+(1+r^{1/2})^2]^{-1}=(2+2r^{1/2}+r)^{-1}$ so that
$r_0=2b$ and $(p,c)=(1,1/\sqrt 2)$ in their parameters.

We note that both $V_1$ and $V_2$ in fact reduce to the incomplete Beta
functions, which may be useful for the purpose of numerical calculations.
Furthermore, if $4p$ is an integer, both $V_1$ and $V_2$ reduce to
elementary functions of $\psi$, and so do the velocity second moments. In
particular, if $p=1$ [for which $\psi=(2+r)^{-1}$],
\begin{equation}
\langle v_r^2\rangle=
\frac12(1+r)(2+r)^3\,\ln\left(\frac{2+r}{1+r}\right)
-\frac{128+303r+250r^2+90r^3+12r^4}{24(2+r)};
\end{equation}
\begin{equation}
\langle v_\mathrm T^2\rangle=
\frac12(1+2r)(2+r)^3\,\ln\left(\frac{2+r}{1+r}\right)
-\frac{64+217r+211r^2+84r^3+12r^4}{12(2+r)}.
\end{equation}

\section{Generalized Plummer Models}
\label{sec:gp}

Next, let us consider the one-parameter family of potential-density pairs,
\begin{equation}
\psi=\frac{GM}{(a^p+r^p)^{1/p}}\,;\qquad
\rho=\frac{(p+1)M}{4\pi}\,\frac{a^p}{r^{2-p}(a^p+r^p)^{2+1/p}},
\label{eq:pdpair}
\end{equation}
which was originally introduced by \citet{Ve79}. This includes the
\citet{He90} model ($p=1$) and the \citet{Pl11} model ($p=2$) as
particular cases. \citet{EA05} recently found that this family can be
constructed from power-law DFs of the form of $f(E,L)\propto
L^{p-2}E^{(3p+1)/2}$. Here we construct slightly more complicated
DFs that build this family of models. We first find every DF of the
form of equation (\ref{eq:ansatz}), and then use them to build models
with a more general variation of anisotropy.

\subsection{Distribution Functions with Constant Anisotropy Parameter}

With $G=M=a=1$, we find that
\begin{equation}
r^{2\beta}\rho
=\frac{p+1}{4\pi}\psi^{p+3-2\beta}(1-\psi^p)^{1-2(1-\beta)/p},
\label{eq:gpd}
\end{equation}
and that
\begin{equation}
\left.\frac{d^m}{d\psi^m}\left(r^{2\beta}\rho\right)\right|_{\psi=0}=0
\end{equation}
for $m<p+3-2\beta$. Then, equation (\ref{eq:disint}) reduces to
\begin{equation}
f(E,L)=
\frac{2^{\beta-1}}{(2\pi)^{5/2}}
\frac{p+1}{\Gamma(n-\beta-1/2)\Gamma(1-\beta)}\,
\frac1{L^{2\beta}} 
\frac{d^{n+1}}{dE^{n+1}}
\int_0^E\frac{\psi^{p+3-2\beta}}{(1-\psi^p)^\lambda}
\frac{d\psi}{(E-\psi)^{3/2-\beta-n}},
\label{eq:gpdf}
\end{equation}
where $n=\lfloor 3/2-\beta\rfloor$ is the integer floor of
$3/2-\beta$, and $\lambda+1=2(1-\beta)/p$. If $\lambda\ge0$, it is at
least formally possible to derive the series expression of the DF for
arbitrary $p$ and $\beta$ from the integral form, namely,
\begin{equation}
f(E,L)=\frac{2^{\beta-1}}{(2\pi)^{5/2}}\frac{p+1}{\Gamma(1-\beta)}\,
\frac{E^{p+3/2-\beta}}{L^{2\beta}}\,\sum_{k=0}^\infty
\frac{\Gamma(pk+p+4-2\beta)}{\Gamma(pk+p+5/2-\beta)}
\frac{\Gamma(k+\lambda)}{\Gamma(\lambda)} \frac{E^{pk}}{k!},
\label{eq:gpdfs}
\end{equation}
which becomes the generalized hypergeometric series if $p$ is a
rational number. On the other hand, if $-1<\lambda<0$, we find that
$\lim_{E\rightarrow1^-}f(E,L)<0$ so that the corresponding DF is
unphysical. Hence, for the potential-density pair of equation
(\ref{eq:pdpair}), the constant anisotropy DF is physical only if
$p\le2(1-\beta)$. This means that the hypervirial models of
\citet{EA05}, which have an anisotropy parameter $\beta=1-(p/2)$, have
the maximally radially biased velocity dispersions for a given $p$ and
a constant $\beta$. Although there exist models with $\beta$ locally
exceeding $1-(p/2)$ for the generalized Plummer sphere of given $p$
\citep[see e.g.,][]{BD02}, $\beta$ at the center cannot be greater than
$1-(p/2)$ for physically valid DFs. This is in agreement with the cusp
slope--central anisotropy theorem derived by \citet{AE05b}.

The corresponding velocity dispersions can be found either from
equation (\ref{eq:vint}) or by solving Jeans equation. The results can
be generally expressed using the incomplete Beta functions. The
equivalent results expressed in terms of hypergeometric functions are
also found in equation (9) of \citet{EA05}.

For particular values of $p$ or $\beta$, the DF of equation
(\ref{eq:gpdf}) (or equivalently, eq.~\ref{eq:gpdfs}) reduces to a
simpler form. For example, if $\beta=1/2$, the DF is (see
eq.~\ref{eq:dishalf})
\begin{equation}
f(E,L)=\frac{p+1}{(2\pi)^3L}\frac{E^{p+1}}{(1-E^p)^{1/p}}
\left[(p+2)-(2p+1)E^p\right],
\end{equation}
where the positive definiteness of the DF implies that $0<p\le1$.
Similarly, for $\beta=-1/2$, the DF is
\begin{equation}
f(E,L)=\frac{(p+1)L}{(2\pi)^3}\frac{E^{p+2}}{(1-E^p)^{1+3/p}}
\left[(p+3)(p+4)-(5p^2+12p+3)E^p+2p(2p+1)E^{2p}\right],
\end{equation}
which is nonnegative for $0\le E\le1$ if $0<p\le3$. On the other
hand, the DF of the Hernquist model with constant $\beta$ reduces to
the hypergeometric function \citep[see e.g.,][]{BD02}
\begin{equation}
f(E,L)=\frac{2^\beta\Gamma(5-2\beta)}
{(2\pi)^{5/2}\Gamma(7/2-\beta)\Gamma(1-\beta)}\
\frac{E^{5/2-\beta}}{L^{2\beta}}\,
\F{1-2\beta}{5-2\beta}{\frac72-\beta}E,
\end{equation}
which is nonnegative everywhere if $\beta\le1/2$, whereas that for
the constant-$\beta$ Plummer model reduces to
\begin{equation}
f(E,L)=\frac{3\cdot2^{\beta-1}\Gamma(6-2\beta)}
{(2\pi)^{5/2}\Gamma(9/2-\beta)\Gamma(1-\beta)}\
\frac{E^{7/2-\beta}}{L^{2\beta}}\,
\mbox{}_3F_2\!\left(-\beta,3-\beta,\frac72-\beta;
\frac{9-2\beta}4,\frac{11-2\beta}4;E^2\right),
\end{equation}
which is physical if $\beta\le0$.

\subsection{Distribution Functions with Outwardly Decreasing
Anisotropy Parameter}

\begin{figure}
\epsscale{0.4}
\plotone{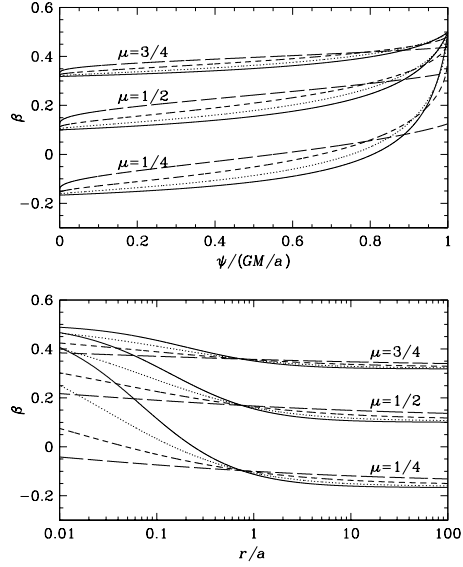}
\caption{\label{fig:aniso} Behavior of the anisotropy parameter
generated by the DFs (eq.~\ref{eq:sdf}) for the
generalized Plummer models. \textit{Solid lines}, $p=1$ (Hernquist model);
\textit{dotted lines}, $p=3/4$; \textit{short-dashed lines}, $p=1/2$;
\textit{long-dashed lines}, $p=1/4$.}
\end{figure}

Since equation (\ref{eq:gpd}) is valid for any constant $\beta$ with
common $\rho$, it is, in general, also possible to express the density
profile of the generalized Plummer model by
\begin{equation}
\rho=\frac{p+1}{4\pi}\sum_i\mu_ir^{-2\beta_i}
\psi^{p+3-2\beta_i}(1-\psi^p)^{1-2(1-\beta_i)/p}\,;\qquad
\sum_i\mu_i=1,
\end{equation}
where the parameters $\mu_i>0$ are normalized weights. Consequently, the
superposition of the constant anisotropy DFs,
$f(E,L)=\sum_i\mu_if_i(E,L)$, still builds the generalized Plummer
sphere. Here each $f_i(E,L)$ is given by equation (\ref{eq:gpdf}) (or
equivalently, eq.~\ref{eq:gpdfs}) in which $\beta$ is replaced by
$\beta_i$. The velocity second moments for these models are found to be
\begin{equation}
\rho\langle v_r^2\rangle=\rho\psi\,\sum_i
\frac{\mu_i\,S\!(r;p,\beta_i)}{p+4-2\beta_i}\,;\qquad
\rho\langle v_\mathrm T^2\rangle=\rho\psi\,\sum_i
\frac{2\mu_i(1-\beta_i)\,S\!(r;p,\beta_i)}{p+4-2\beta_i},
\end{equation}
where
\begin{equation}
S\!(r;p,\beta)=
\F{\frac{2-2\beta}{p}-1}1{\frac{4-2\beta}p+2}{-\frac1{r^p}}
=\left(r\psi\right)^p\F{\frac2p+3}1{\frac{4-2\beta}p+2}{\psi^p},
\label{eq:hypg}
\end{equation}
and therefore the anisotropy parameter $\beta$ is no longer constant.

For the particular case in which the sum only contains two terms (such
that $\mu_1=\mu$, $\mu_2=1-\mu$, and $\mu_i=0$ for all other $i$ values),
the anisotropy parameter is
\begin{equation}
\beta=
\frac{\beta_1\mu(p+4-2\beta_2)\,S\!(r;p,\beta_1)
+\beta_2(1-\mu)(p+4-2\beta_1)\,S\!(r;p,\beta_2)}
{\mu(p+4-2\beta_2)\,S\!(r;p,\beta_1)
+(1-\mu)(p+4-2\beta_1)\,S\!(r;p,\beta_2)},
\end{equation}
which is a decreasing function of $r$, provided that $0<\mu<1$ and
$\beta_1\ne\beta_2$. If we assume $\beta_2<\beta_1\le1-(p/2)$ and
$0<\mu<1$, the limiting values are found to be
\begin{equation}
\lim_{r\rightarrow\infty}\beta=
\frac{\beta_1\mu(p+4-2\beta_2)+\beta_2(1-\mu)(p+4-2\beta_1)}
{\mu(p+4-2\beta_2)+(1-\mu)(p+4-2\beta_1)}.
\end{equation}
On the other hand,
\begin{equation}
\beta(r=0)=\beta_1\qquad\mbox{if }\ \beta_1\ge1-p,
\end{equation}
or
\begin{equation}
\beta(r=0)=
\frac{\beta_1\mu(1-p-\beta_2)+\beta_2(1-\mu)(1-p-\beta_1)}
{\mu(1-p-\beta_2)+(1-\mu)(1-p-\beta_1)}
\qquad\mbox{if }\ \beta_1\le1-p.
\end{equation}
The simplest example of the DFs of this kind is obtained when we
choose $\beta_1=1/2$ and $\beta_2=-1/2$:
\begin{equation}
f(E,L)=
\frac{p+1}{(2\pi)^3}\left[\mu\frac{E^{p+1}}L\,
\frac{(p+2)-(2p+1)E^p}{(1-E^p)^{1/p}}
+
(1-\mu)LE^{p+2}\,
\frac{(p+3)(p+4)-(5p^2+12p+3)E^p+2p(2p+1)E^{2p}}{(1-E^p)^{1+3/p}}\right],
\label{eq:sdf}
\end{equation}
which is physical if $p\le1$. The behavior of the anisotropy parameter
$\beta$ for the model given by DFs of equation (\ref{eq:sdf}) is shown
in Figure~\ref{fig:aniso}. Again we note that equation (\ref{eq:hypg})
in general reduces to the incomplete Beta function while, for this case,
it further reduces to an expression involving only elementary functions of
$\psi$ if $2p$ is an integer, and so and so do $\beta$ and the velocity
second moments.

\section{An Application: The Mass of the Milky Way}
\label{sec:app}

\begin{figure}
\epsscale{0.5}
\plotone{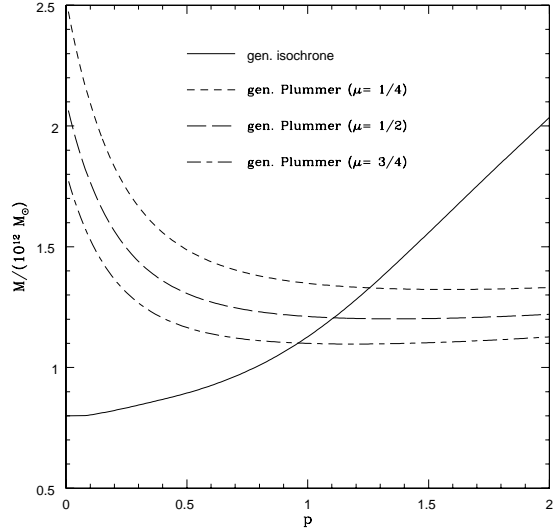}
\caption{\label{fig:massaniso} Curves illustrating the degeneracy
between mass and anisotropy. All the models have the same half-mass
radius and reproduce the same mean kinetic energy $\langle T_{rr}
\rangle$ of the satellite population of the Milky Way. The inferred
mass of the Milky Way (in units of $10^{12}\ M_\sun$) is
plotted against the parameter $p$, which controls the velocity
anisotropy.}
\end{figure}

As an application of the preceding theory, let us consider the problem
of estimating the mass of the Milky Way from the radial
velocities of its satellites. There are 27 distant globular clusters
and dwarf galaxies at Galactocentric distance $r$ greater than 20 kpc
\citep[see e.g., tables 2 and 3 of][]{WE99}. These range from Arp~2 at
$r=20\ \mbox{kpc}$ out to Leo~I at $r=254\ \mbox{kpc}$. The median
distance of the satellite population is $r_\mathrm{med}\approx 38\
\mbox{kpc}$. The number density of the satellites falls off roughly
like $r^{-3.5}$. This dataset is often used to estimate the mass of
the Milky Way \citep{Ko96,WE99}, as the \ion H1 rotation curve
cannot be traced beyond $\sim$20 kpc. The satellites are therefore the
only available probe of the gravitational field of the distant halo.

All the satellites have accurate radial velocities, but proper motions
are available only for a handful of the closest objects. Of course,
the heliocentric radial velocity and the Galactocentric radial
velocity are basically the same (once the motion of the Sun has been
taken out), as the Galactocentric distance of the Sun is much smaller
than those of the satellites. The kinematic data therefore constrains the mean
kinetic energy in the radial direction $T_{rr}=\langle v_r^2\rangle$.
Again using the data in tables 2 and 3 of \citet{WE99}, we find that
$T_{rr}\approx(129\ \mbox{km s$^{-1}$})^2$.

To model the Milky Way halo, we assume that its potential and density
take the form of either the generalized isochrone or the generalized
Plummer models. Furthermore, we assume that the number density of
satellites shadows the total density \citep[e.g.,][]{Ko96,WE99}. This
means that the satellites are drawn from DFs of the form given
by either equation (\ref{eq:diffiso}) or equation (\ref{eq:sdf}). We set the
scale length so that the half-mass radius of the assumed density
profile corresponds to the observed median radius of the population.
The mass of the Milky Way $M$ enters into the potential and the
DF. It can therefore be constrained by requiring that the mean kinetic
energy in the radial direction be consistent with the data, viz,
\begin{equation} 
T_{rr}=
\frac{\int\!\rho\langle v_r^2\rangle\,d^3\!\bm r}
{\int\!\rho\,d^3\!\bm r}
\approx16,640\ \mbox{km$^2$ s$^{-2}$}.
\label{eq:crux}
\end{equation}
The parameter $p$ in the DFs controls the variation of the velocity
anisotropy with radius. Hence, equation (\ref{eq:crux}) defines a line
in the plane $(M,p)$, which is the \emph{mass-anisotropy degeneracy
curve} for the model. Figure~\ref{fig:massaniso}, inferred from
numerical integrations, shows such curves computed for a number of the
generalized isochrones and the generalized Plummer models.
 
In this problem, the mean kinetic energy in the radial direction is
fixed. Models with tangential anisotropy at large radii therefore lead
to a higher inferred mass for the Milky Way galaxy. The relative
location of the curves in Figure~\ref{fig:massaniso} can be understood
by examining the behavior of the anisotropy parameter for the models.
More importantly, as the velocity anisotropy changes, the inferred
mass of the Milky Way ranges from $0.8\times10^{12}\ M_\sun$ to
$2.5\times10^{12}\ M_\sun$. Now, the velocity anisotropy of the
satellite galaxy population is unknown, so the mass-anisotropy
degeneracy \emph{by itself} enforces a factor of $\sim$3 uncertainty in the
mass of the Milky Way. This fundamental limitation can only be
overcome by proper-motion data.

Of course, there are other uncertainties; for example, the half-mass
radius of the satellite population used to set the scale length in
these calculations is also not well known (note that the locations of
the satellites are likely to be biased toward their apogalacticon)
and has an uncertainty of
$\sim$20\%. This carries over into an additional $\sim$20\%
uncertainty in the inferred mass. This is small in comparison with the
mass-anisotropy degeneracy, but not insignificantly so.

\section{Conclusions}
\label{sec:conc}

This paper provides a number of flexible and simple galaxy models
extended from two classical cored profiles: the isochrone sphere and
Plummer model. We also derive analytic distribution functions (DFs) 
with varying velocity anisotropy for all the models including new DFs
for the classical isochrone sphere and the Hernquist model. 

DFs of the form $f(E,L)=L^{-2\beta}f_E(E)$ build a spherical system
with a constant anisotropy parameter $\beta$, where $E$ is the binding
energy and $L$ is the specific angular momentum. The assumption of constant
anisotropy is not enough to provide realistic models, as both
observational data and numerical simulations suggest that dark halos
and their constituents (such as satellites) have velocity anisotropies
that vary with radius \citep[e.g.,][]{Wh85,vdM94,HM05}.

Here we have explored models whose DFs are superpositions of two or
more such terms, namely,
\begin{equation}
f(E,L) = \sum_iL^{-2\beta_i}f_i(E).
\end{equation}
This can be done in two ways. First, as illustrated by the generalized
isochrones in \S~\ref{sec:gis}, we can sometimes find a splitting of the density
in the form
\begin{equation}
\rho(r) = \sum_ir^{-2\beta_i}g_i(\psi(r)).
\end{equation}
and apply the inversion separately to each component. For the
generalized isochrones, this provides models with either decreasing
anisotropy parameter $\beta$ in the inner region and increasing $\beta$ in
the outer parts, or decreasing $\beta$ throughout. Second, as
illustrated by the generalized Plummer models in \S~\ref{sec:gp}, we can always
add together weighted sums of constant anisotropy DFs, each of which
individually reproduces the required density. This appears to provide
models with decreasing anisotropy parameter $\beta$ throughout. At
large radii, the DF is dominated by the component of smallest $\beta$,
and the models can be either radially or tangentially anisotropic
in the outer parts, according to choice. For both cases, it appears
that the anisotropy parameter always decreases on beginning to move
outward from the very center.

There have been a number of previous algorithms suggested for building
anisotropic DFs with varying anisotropy. For
example, the Osipkov-Merritt algorithm \citep{Os79,Me85}
builds models that are isotropic in the inner parts and tend to
extreme radial anisotropy ($\beta \rightarrow 1$) in the outer
parts. \citet{Ge91} provided a clever algorithm for building separable
DFs of the form
\begin{equation}
f=g(E)h(L/L_\mathrm c(E)),
\end{equation}
where $L_\mathrm c$ is the angular momentum of a circular orbit
with energy $E$. The analytic examples provided by \citet{Ge91} all
have increasing radial anisotropy in the outer parts. His method can
be used to generate tangentially anisotropic models, but usually at
the cost of a numerical inversion. The methods developed in this paper
therefore provide a complement to the existing inversions.

Tangential anisotropy often arises in accreted populations, such as
satellite galaxies. As an application of our models, we have shown
that the mass-anisotropy degeneracy by itself provides a factor
$\sim$3 uncertainty in the mass of the Milky Way galaxy as deduced
from the kinematics of its satellites.

The density profiles of models studied in this paper typically
fall off faster than
$\rho\sim r^{-3}$ and slower than $\rho\sim r^{-5}$ at large radii,
which are in good agreement with both observations of spheroidal
components of galaxies and results from numerical simulations. At
small radii, they are typically cusped, with cusp indices in the range
suggested by numerical simulations. In particular, there are members
with density cusps like $\rho\sim r^{-1}$, $\sim r^{-4/3}$, and
$\sim r^{-3/2}$, which have been suggested as important on
cosmogonic grounds \citep{NFW95,EC97,Mo98}, although observational
evidence seems to favor the presence of a constant-density core
\citep{TKA98,PW00,dB01,KW03,DGS04}. However, even if real galaxies
are cored, the core would be such a small fraction of the halo that it
is not practically important for studies of the whole. Our families of
new models therefore should find widespread application in the
modeling of galaxies and dark halos.

\acknowledgements
We thank C. Hunter,
who made a number of interesting comments on the various incarnations
of this paper.

\begin{appendix}

\section{Constant Anisotropy Distribution Functions}
\label{sec:con}

In this appendix we summarize some results regarding
constant-anisotropy DFs used in the main body of the paper. For the DF given in
equation (\ref{eq:ansatz}), direct integration over velocity space
gives
\begin{equation}
\rho=r^{-2\beta}
\frac{(2\pi)^{3/2}\Gamma(1-\beta)}{2^\beta\Gamma(3/2-\beta)}
\int_0^\psi\!(\psi-E)^{1/2-\beta}f_E(E)\,dE.
\label{eq:den}
\end{equation}
The formula for inversion for the unknown function $f_E(E)$ is given
by \citep{Cu91}
%
\begin{eqnarray}
f(E,L)&=&
\frac1{L^{2\beta}}
\frac{2^\beta}{(2\pi)^{3/2}\Gamma(1-\alpha)\Gamma(1-\beta)}\
\frac d{dE}\int_0^E\frac{d\psi}{(E-\psi)^\alpha}\frac{d^nh}{d\psi^n}
\nonumber\\&=&
\frac1{L^{2\beta}}
\frac{2^\beta}{(2\pi)^{3/2}\Gamma(1-\alpha)\Gamma(1-\beta)}
\left[\int_0^E\frac{d\psi}{(E-\psi)^\alpha}\frac{d^{n+1}h}{d\psi^{n+1}}
+\frac1{E^\alpha}\left.\frac{d^nh}{d\psi^n}\right|_{\psi=0}\right],
\label{eq:disint}
\end{eqnarray}
%
where $h(\psi)=r^{2\beta}\rho$ is expressed as a function of $\psi$,
and $n=\lfloor 3/2-\beta\rfloor$ and $\alpha=(3/2-\beta)-n$ are the
integer floor and the fractional part of $3/2-\beta$, respectively.
This includes
equation (\ref{eq:eddington}) as a particular case of $\beta=0$ ($n=1$
\& $\alpha=1/2$). If $\beta$ is a half-integer constant (i.e.,
$\beta=$ $1/2$, $-1/2$, $-3/2$, and so on), the above inversion
simplifies to
\begin{equation}
f(E,L)=\frac1{2\pi^2L^{2\beta}}\frac1{(-2\beta)!!}\
\left.\frac{d^{3/2-\beta}h}{d\psi^{3/2-\beta}}\right|_{\psi=E},
\label{eq:dishalf}
\end{equation}
that is, this only involves differentiations in the process
\citep{Cu91}. In addition, the expression for the differential energy
distribution (DED) is found to be
\begin{equation}
\left.\frac{d\rho}{dE}\right|_{E=E_0}
=\int\frac{f_E(E)}{L^{2\beta}}\delta(E-E_0)\,d^3\!\bm v
=\frac{(2\pi)^{3/2}\Gamma(1-\beta)}{2^\beta\Gamma(3/2-\beta)}
\frac{(\psi-E_0)^{1/2-\beta}}{r^{2\beta}}\,f_E(E_0)\,\Theta(\psi-E_0)\,;
\end{equation}
\begin{equation}
\frac{dM}{dE}=\int\frac{d\rho}{dE}\,d^3\!\bm r
=f_E(E)\,\frac{(2\pi)^{5/2}\Gamma(1-\beta)}{2^{\beta-1}\Gamma(3/2-\beta)}
\int_0^{r_E}(\psi-E)^{1/2-\beta}r^{2(1-\beta)}\,dr,
\label{eq:ded}
\end{equation}
where $\psi(r_E)=E$, and $\delta(x)$ and $\Theta(x)$ are the Dirac
delta `function' and the Heaviside unit step function, respectively.
In particular, if $\beta=1/2$, the last integral does not explicitly
involve the potential, and so the DED may be found simply as
$dM/dE=4\pi^3r_E^2f_E(E)$.

It is also straightforward to find the integral for the second
velocity moments;
\begin{eqnarray}
\rho\langle v_r^2\rangle
&=&\frac{(2\pi)^{3/2}}{2^\beta r^{2\beta}}
\frac{\Gamma(1-\beta)}{\Gamma(5/2-\beta)}
\int_0^\psi\!(\psi-E)^{3/2-\beta}f_E(E)dE\,;
\label{eq:diffy}\\
\rho\langle v_\mathrm T^2\rangle
&=&\frac{2(2\pi)^{3/2}}{2^\beta r^{2\beta}}
\frac{\Gamma(2-\beta)}{\Gamma(5/2-\beta)}
\int_0^\psi\!(\psi-E)^{3/2-\beta}f_E(E)dE
=2(1-\beta)\rho\langle v_r^2\rangle.
\end{eqnarray}
In fact, the velocity dispersion of any constant anisotropy models may
be found from $h(\psi)$ without explicitly evaluating the DF
\citep{De87,BD02}. On differentiating equation (\ref{eq:diffy}) with
respect to $\psi$ and noting that the right-hand side is the same as
equation (\ref{eq:den}), that is, $\partial(\rho\langle v_r^2\rangle)/
\partial\psi=\rho$, it follows that
\begin{equation}
\langle v_r^2\rangle
=\frac1{\rho}\int_0^\psi\!d\psi'\,\rho(r,\psi')
=\frac1{h(\psi)}\int_0^\psi\!d\psi'\,h(\psi')
=\frac1{r^{2\beta}\rho}
\int_\infty^r\!dr'\frac{d\psi'}{dr'}\,{r'}^{2\beta}\rho.
\label{eq:vint}
\end{equation}
We note that equation (\ref{eq:vint}) can also be derived from the
integral solution of Jeans' equation with a constant $\beta$
\citep[see eq.~9 of][]{EA05}. This also indicates that equation
(\ref{eq:vint}) is actually a general result for a constant $\beta$
model, regardless of the specific form of the DF that generates the
model.

As an example, let us think of the DFs that build the generalized
isochrone sphere with a constant anisotropy parameter.
First, the DF and the DED for the generalized isochrone sphere with
$\beta=1/2$ can be found from equation~(\ref{eq:dishalf});
\begin{equation}
f(E,L)=\frac1{(2\pi)^3L}
\frac{E^2\,g(E)}{(1-E)^{2p}\left[(1-E)^p-E^p\right]^{1/p}}
\,;\qquad
\frac{dM}{dE}=
\frac{\left[(1-E)^p-E^p\right]^{1/p}}{2(1-E)^{2p}}\
g(E)
\end{equation}
\begin{equation}
g(E)=
E^{2p-1}\left[6(p+1)E-6E^2-(p+1)(2p+1)\right]
+
(1-E)^pE^{p-1}\left[(p+1)(p+2)+12E^2-6(p+2)E\right]
+6(1-E)^{2p+1}.
\end{equation}
Here, because $0\le E\le\psi\le1/2$, the nonnegativity of the DF is
satisfied only if $0<p\le1$. In addition, the integration involved in
the inversion (eq.~\ref{eq:disint}) is also analytically tractable if
$\beta=1-(p/2)$. That is, for the density profile of the generalized
isochrone sphere, we find that
\begin{equation}
r^{2-p}\rho=\frac1{2\pi}\psi^{p+2}(1-\psi)^{1-p}
+\frac1{8\pi}\frac d{d\psi}\left[\psi^{2p+2}(1-\psi)^{2-2p}\right].
\end{equation}
This leads to the DF of the form of equation (\ref{eq:ansatz})
\begin{eqnarray}
f(E,L)&=&\frac1{L^{2-p}}\frac1{2^{p/2+3/2}\pi^{5/2}\Gamma(p/2)}
\nonumber\\&\times&
\left[\frac{\Gamma(2p+3)}{4\Gamma(3p/2+3/2)}E^{(3p+1)/2}
\F{2p-2}{2p+3}{\frac{3p+3}2}E
+
\frac{\Gamma(p+3)}{\Gamma(p/2+5/2)}E^{(p+3)/2}
\F{p-1}{p+3}{\frac{p+5}2}E\right],\qquad
\end{eqnarray}
which build a constant anisotropy model with $\beta=1-(p/2)$.

If $p=1$ (corresponding to the Hernquist model) and $\beta=1/2$, the
DF and the DED further reduce to a particularly simple form;
\begin{equation}
f(E,L)=\frac3{2\pi^3}\frac{E^2}L\,;\qquad
\frac{dM}{dE}=6(1-2E)^2,
\end{equation}
which was originally found by \citet{BD02} for the \citet{He90}
model. For $p=2$ and $\beta=0$, the well-known expression for the DF
of the isotropic isochrone sphere \citep[see e.g.,][p. 239]{He60,BT87}
is recovered;
\begin{equation}
f(E)=\frac{1}{2^{15/2}\pi^3(1-E)^4}
\left[(16E^2+28E-9)\,\frac{3\arcsin\!\sqrt{E}}{\sqrt{1-E}}
+\left(27-66E+320E^2-240E^3+64E^4\right)\sqrt{E}\right].
\end{equation}
%

Finally, let us think of the DED of the generalized Plummer models
corresponding to the DF given in equation~(\ref{eq:gpdf}). For this
case, equation~(\ref{eq:ded}) reduces to a similar integral as 
equation~(\ref{eq:gpdf})
\begin{equation}
\frac{dM}{dE}
=\frac{(2\pi)^{5/2}\Gamma(1-\beta)f_E(E)}{2^{\beta-1}}
\frac{1}{\Gamma(3/2-\beta)}
\int_E^1\frac{(1-\psi^p)^{(3-2\beta)/p-1}}{\psi^{2(2-\beta)}}
(\psi-E)^{1/2-\beta}\,d\psi.
\label{eq:dedgp}
\end{equation}
Like the DF, equation~(\ref{eq:dedgp}) reduces to a closed form
function of $E$ for particular values of $p$ or $\beta$. Among them
are for $\beta=1/2$
\begin{equation}
\frac{dM}{dE}=f_E(E)\,\frac{(2\pi)^3}{2}\frac{(1-E^p)^{2/p}}{E^2}
=\frac{p+1}{2}\,(1-E^p)^{1/p}\left[(p+2)-(2p+1)E^p\right]E^{p-1},
\label{eq:dedpp}
\end{equation}
for $\beta=-1/2$,
\begin{eqnarray}
\frac{dM}{dE}&=&f_E(E)\,\frac{(2\pi)^3}{4(p+4)}\frac{(1-E^p)^{1+4/p}}{E^3}
\F{1}{\frac{1}{p}+1}{\frac{4}{p}+2}{1-E^p}
\nonumber\\&=&\frac{p+1}{p+4}\frac{(1-E^p)^{1/p}}{4}
\left[(p+3)(p+4)-(5p^2+12p+3)E^p+2p(2p+1)E^{2p}\right]E^{p-1}
\F{1}{\frac{1}{p}+1}{\frac{4}{p}+2}{1-E^p},\qquad
\label{eq:dedpn}
\end{eqnarray}
and for $p=1$,
\begin{eqnarray}
\frac{dM}{dE}&=&f_E(E)\,
\frac{(2\pi)^{5/2}\Gamma(3-2\beta)\Gamma(1-\beta)}
{2^{\beta-1}\Gamma(9/2-3\beta)}
\frac{(1-E)^{7/2-3\beta}}{E^{5/2-\beta}}
\F{\frac{1}{2}-\beta}{\frac{3}{2}-\beta}{\frac{9}{2}-3\beta}{1-E}
\nonumber\\&=&
\frac{2\Gamma(5-2\beta)\Gamma(3-2\beta)}{\Gamma(9/2-3\beta)\Gamma(7/2-\beta)}
(1-E)^{7/2-3\beta}
\F{1-2\beta}{5-2\beta}{\frac{7}{2}-\beta}{E}
\F{\frac{1}{2}-\beta}{\frac{3}{2}-\beta}{\frac{9}{2}-3\beta}{1-E}.
\end{eqnarray}
This suggests that $dM/dE$ diverges as $E\rightarrow0$ for $p<1$, that
it vanishes for $p>1$, and that it has a finite limiting value of
$8(2-\beta)/(5-2\beta)$ for $p=1$. We believe that this is because the
behavior of $dM/dE$ near $E=0$ is dominated by the stars at large
radii and the $r^{-4}$ fall-off marks a critical point \citep{EA06}.

\end{appendix}

\end{document}